\documentclass[a4paper,11pt]{article}
\pdfoutput=1
\usepackage{wrapfig}
\usepackage[dvipsnames]{xcolor}
\usepackage{setspace}
\PassOptionsToPackage{hypertexnames=false}{hyperref}
\usepackage{jcappub}
\usepackage[T1]{fontenc}
\usepackage{multirow}
\usepackage{float}

\title{\boldmath Informative Priors on Primordial Non-Gaussianity Bias $b_{\phi}$ From Galaxy Formation}

\author[a,1]{Anne Moore,\note{Corresponding author.}}
\author[b,c]{Lucia A. Perez,}
\author[d,a]{Elisabeth Krause}

\affiliation[a]{Department of Physics, University of Arizona, Tucson, AZ 85721, USA}
\affiliation[b]{Center for Computational Astrophysics, Flatiron Institute, 162 5th Ave, New York, NY 10010, USA}
\affiliation[c]{Department of Astrophysical Sciences,
Princeton University, 4 Ivy Lane, Princeton, NJ 08544, USA}
\affiliation[d]{Steward Observatory, University of Arizona, 922 N Cherry Ave, Tuscon, AZ 85719, USA}

\emailAdd{amoore8@arizona.edu}
\emailAdd{lperez@flatironinstitute.org}
\emailAdd{krausee@arizona.edu}

\abstract{Constraining primordial non-Gaussianity via its scale-dependent imprint on galaxy clustering requires knowledge of the bias parameter $b_{\phi}$, which is exactly degenerate with $f^{\rm{loc}}_{\rm{NL}}$ at leading order. To break this degeneracy, current analyses adopt the relation $\left(b_{\phi} = 2\delta_c\left(b_1 - 1\right)\right)$ based on the assumption of a universal mass function. This relation is known to break down for physically motivated galaxy selections, introducing systematic errors in the inferred $f^{\rm{loc}}_{\rm{NL}}$ that scale directly with the assumed $b_{\phi}$ prior. We present a framework to construct physically motivated, observation-conditioned priors on $b_{\phi}$ by marginalizing over galaxy formation uncertainties. We use the CAMELS-SAM simulation suite, augmented by separate Universe simulations, to measure galaxy formation observables, like the stellar mass function (SMF) and the stellar-to-halo mass relationship (SHMR), and $b_{\phi}$ across a range of galaxy formation parameters. From these measurements, we construct a distribution of $b_{\phi}$ conditioned on observations, and we select our galaxy sample to resemble the DESI Emission Line Galaxy (ELG) sample. Conditioning on the SMF or SHMR decreases $\sigma_{b_{\phi}}$ from $0.69$ to $0.08$ and $0.02$ respectively -- reductions of $88\%$ and $97\%$ -- with consistent results when conditioning on the observed data directly. Despite substantial shifts in the galaxy formation posteriors driven by known SC-SAM discrepancies at high halo masses, the resulting $b_{\phi}$ distributions remain mutually consistent across all observables. The SMF and SHMR are found to carry sufficient constraining power to reduce the galaxy formation uncertainty in $b_{\phi}$ relevant for $f^{\rm{loc}}_{\rm{NL}}$ inference with next-generation spectroscopic surveys}

\begin{document}
\maketitle
\flushbottom

\section{Introduction}\label{sec:Intro} 
Constraining the primordial density fluctuations generated during inflation is one of the central open problems in modern cosmology. Different inflationary models predict distinct observational signatures in the primordial density field, testable through measurements of both the Cosmic Microwave Background (CMB) and large-scale structure (LSS). 

CMB temperature and polarization power spectra constrain the scalar spectral index $n_{\mathrm{s}}$, which characterizes the scale dependence of the primordial curvature power spectrum. The tightest current constraints come from Planck~\cite{2020}, yielding $n_{\mathrm{s}} = 0.9649 \pm 0.0042$, consistent with scale-invariance breaking as predicted by inflationary models. Tensor perturbations in the primordial field produce B-mode polarization in the CMB through primordial gravitational waves; measurements of the tensor-to-scalar ratio $r$ constrain the energy scale of inflation and discriminate between inflationary scenarios. Planck~\cite{2020} constrains $r_{0.002} < 0.056$, placing an upper bound on the inflationary energy scale. 

Beyond the primordial power spectrum, inflationary models can be further differentiated through higher-order statistics of the primordial density field, as many models predict non-Gaussian curvature perturbations. For local-type primordial non-Gaussianity (PNG), perturbations to the primordial gravitational potential $\phi\left(\boldsymbol{x}\right)$ are parametrized as~\cite{Komatsu_2001}  

\begin{equation} \phi\left(\boldsymbol{x}\right) = \phi_{\rm{G}}\left(\boldsymbol{x}\right) + f^{\rm{loc}}_{\rm{NL}}\lbrack \phi_{\rm{G}}\left(\boldsymbol{x}\right)^2 - \langle\phi_{\rm{G}}\left(\boldsymbol{x}\right)^2\rangle\rbrack 
\end{equation}  
where $\phi_G\left(x\right)$ is a Gaussian random field, $\langle\cdots\rangle$ denotes the ensemble average, and $f^{\rm{loc}}_{\rm{NL}}$ quantifies the deviation from Gaussianity. Single-field inflationary models~\cite{Maldacena_2003,Paolo_Creminelli_2004,Creminelli_2011,Tanaka_2011,Pajer_2013} predict small deviations ($\mathcal{O}\left(10^{-2}\right)$), while multi-field models~\cite{alvarez2014testinginflationlargescale,Biagetti_2019,achúcarro2022inflationtheoryobservations} predict larger amplitudes ($f^{\rm{loc}}_{\rm{NL}}\ge 1$). Within the CMB, the PNG signal is constrained through the bispectrum of curvature perturbations: $f^{\rm{loc}}_{\rm{NL}}$ governs the amplitude of fluctuations, while the shape of the momentum triangle encodes the specific generation mechanism. A detection of a local-shape primordial bispectrum signal, which peaks in the squeezed limit ($k_1 \ll k_2,\ k_3$), would rule out the entire class of single-field inflationary models.

Current CMB constraints on $f_{\rm{NL}}^{\rm{loc}}$ are insufficient to discriminate between inflationary models. The tightest bound from Planck yields $f^{\rm{loc}}_{\rm{NL}} = -0.9 \pm 5.1$~\cite{planckcollaboration2019planck2018resultsix}, and future CMB experiments will not substantially improve this result due to cosmic variance~\cite{Baumann_2009}. Reaching the sensitivity required to distinguish between inflationary scenarios will require complementary probes from galaxy clustering.

Large-scale structure surveys access a larger number of large-scale modes than the CMB through their three-dimensional maps of the galaxy distribution, making them well-suited probes of local-type PNG. Non-zero $f^{\rm{loc}}_{\rm{NL}}$ imprints a scale-dependent signature on the galaxy distribution that is most prominent at large scales\cite{PhysRevD.77.123514}. The tightest current LSS constraint is $f^{\rm{loc}}_{\rm{NL}}=-0.0\pm 4.1$  from~\cite{chudaykin2025reanalyzingdesidr13}, which exploits the full DESI DR1 dataset combining the galaxy power spectrum and bispectrum.
Forthcoming analyses from the DESI collaboration~\cite{desicollaboration2016desiexperimentisciencetargeting} and surveys including Euclid~\cite{2018} and SPHEREx~\cite{doré2015cosmologyspherexallskyspectral,Bock_2026} are expected to improve upon these constraints through their larger survey volumes. SPHEREx in particular, is projected to achieve $\sigma\left(f^{\rm{loc}}_{\rm{NL}}\right) \sim 1$ by combining power spectrum and bispectrum measurements, meeting the threshold required to rule out single-field inflationary models.  

The imprint of local PNG on the galaxy power spectrum enters through an additional term in the bias expansion. To leading order, the galaxy density contrast in the presence of local PNG is given by~\cite{McDonald_2008,Slosar_2008,Giannantonio_2010,Baldauf_2011,Assassi_2015}

\begin{equation}\label{biasExpansion}
    \delta_g\left(\textbf{k},z\right) \overset{\rm{LO}}{=} b_1\left(z\right)\delta_m\left(\textbf{k},z\right) + b_{\phi}\left(z\right)f^{\rm{loc}}_{\rm{NL}}\phi\left(\textbf{k},z\right) + \epsilon\left(z\right)
\end{equation}  
where $\delta_m$ is the matter density contrast, $b_1$ is the linear galaxy bias, $b_{\phi}$ describes the response of galaxy number counts to long-wavelength primordial perturbations, and $\epsilon\left(z\right)$ represents the stochastic contribution to the density field. Equation~\ref{biasExpansion} makes explicit the central challenge for $f^{\rm{loc}}_{\rm{NL}}$ constraints from LSS: the exact parameter degeneracy between $f_{\rm{NL}}$ and $b_{\phi}$ at the field level. Absent independent constraints on $b_{\phi}$, the $f^{\rm{loc}}_{\rm{NL}}$ contribution to the scale-dependent bias signal cannot be unambiguously isolated from the galaxy power spectrum. 

Combined constraints from the galaxy power spectrum and bispectrum offer one avenue to break this degeneracy. A term proportional to $b_1^3 f^{\rm{loc}}_{\rm{NL}}$ in the galaxy bispectrum could in principle constrain $f^{\rm{loc}}_{\rm{NL}}$ while marginalizing over the remaining galaxy bias parameters. However, Barreira~\cite{Barreira_2022_ps_bis} demonstrated that large uncertainties on the marginalized terms substantially negate the constraining power of this contribution.  

The standard approach in LSS analyses has instead been to adopt the universality relation~\cite{chaussidon2025constrainingprimordialnongaussianitydesi, Ross_2012, Ho_2015, Castorina_2019, mueller2021clusteringgalaxiescompletedsdssiv}, which follows from assuming a universal halo mass function, to express $b_{\phi}$ as a function of $b_1$~\cite{PhysRevD.77.123514,Slosar_2008,Giannantonio_2010,Matarrese_2008}  

\begin{equation}\label{eq:universality}
    b_{\phi} = 2\delta_c\left(b_1 - 1 \right)\ ,
\end{equation}  

where $\delta_c = 1.686$ is the linear collapse threshold. Under eq.~\ref{eq:universality}, the large-scale power spectrum contribution scales as $\propto \left(b_1 - 1\right)f^{\rm{loc}}_{\rm{NL}}/k^2$, where $b_1$ can be independently constrained from the small-scale power spectrum, thereby isolating the $f^{\rm{loc}}_{\rm{NL}}$ signal. However, the universality relation is known to break down for a range of halo masses and mass definitions~\cite{Grossi_2009,Desjacques_2009,Pillepich_2009,Baldauf_2016,Biagetti_2017,Barreira_2020}. Barreira~\cite{Barreira_2022_bphi} found that it systematically mispredicts $b_{\phi}$ at large $b_1$ values, and IllustrisTNG-based analyses~\cite{Barreira_2022_bphi, Barreira_2020} demonstrated that the relation can vary significantly for galaxy samples selected by stellar mass, total mass, color, or black hole accretion rate.  

Systematic uncertainties in the $b_{\phi}$ model propagate directly into inferred constraints on $f^{\rm{loc}}_{\rm{NL}}$, introducing both precision loss and potential bias in the recovered parameter. Barreira~\cite{Barreira_2022_ps_bis} showed using mock data that the inferred uncertainty on $f^{\rm{loc}}_{\rm{NL}}$ depends sensitively on the assumed $b_{\phi}$ prior; wide, uninformative priors generate projection effects that produce apparent biases in the marginalized $f^{\rm{loc}}_{\rm{NL}}$ constraints. For next-generation surveys to deliver reliable constraints on $f^{\rm{loc}}_{\rm{NL}}$, it is therefore necessary to establish realistic, informative priors on $b_{\phi}$.

Past works have employed simulation-based approaches to place priors on standard one-loop bias parameters and counter terms to sharpen the constraining power of cosmological analyses. For example,~\cite{Shiferaw_2025} calibrated the relation between different galaxy bias parameters from galaxy clustering at the field-level using semi-analytic and hydrodynamic galaxy models. Related work in~\cite{ivanov2024fullshapeanalysissimulationbasedpriors,ivanov2025fullshapeanalysissimulationbasedpriors,ivanov2024millenniumastridgalaxieseffective} measured the distributions of one-loop parameters constructed from a large set of mock catalogs. Notably, all of these approaches condition the galaxy bias priors on small-scale clustering and account for galaxy formation only through the selection of galaxy samples, while the current work explicitly conditions $b_\phi$ on observable properties of galaxy samples to account for astrophysical assembly biases.

Perez et al.~\cite{perez2026impactgalaxyformationgalaxy} demonstrated that both observable galaxy properties --- including the stellar mass function and the stellar metallicity-stellar mass relationship --- and $b_{\phi}$ exhibit substantial variation across simulations with differing galaxy formation realizations. The present work addresses this by constructing a physically motivated prior on $b_{\phi}$ conditioned on galaxy formation observables other than galaxy clustering, marginalizing over the uncertainty in the underlying galaxy formation model. 

Quantitatively, we compute the galaxy-formation marginalized $b_{\phi}$ prior as 
\begin{equation}\label{eqn1:bphiPrior}
    p\left(b_{\phi}|\{Obs\} \right) = \int d \boldsymbol{\theta}_{\rm{gf}}\ p\left(b_{\phi}|\boldsymbol{\theta}_{\rm{gf}} \right) p\left(\boldsymbol{\theta}_{\rm{gf}}|\{Obs\} \right). 
\end{equation} 
Here $\{Obs\}$ denotes the set of observed properties of the galaxy sample and $\boldsymbol{\theta}_{\rm{gf}}$ denotes the galaxy formation parameter space, which is sampled by different simulation realizations. Marginalizing over this parameter vector and weighting each realization by the agreement between its predicted and observed galaxy properties yields a $b_{\phi}$ distribution that incorporates uncertainties in the galaxy formation physics.

\begin{figure}[h] 
\centering 
    \includegraphics[width=1.0\textwidth,height=12cm]{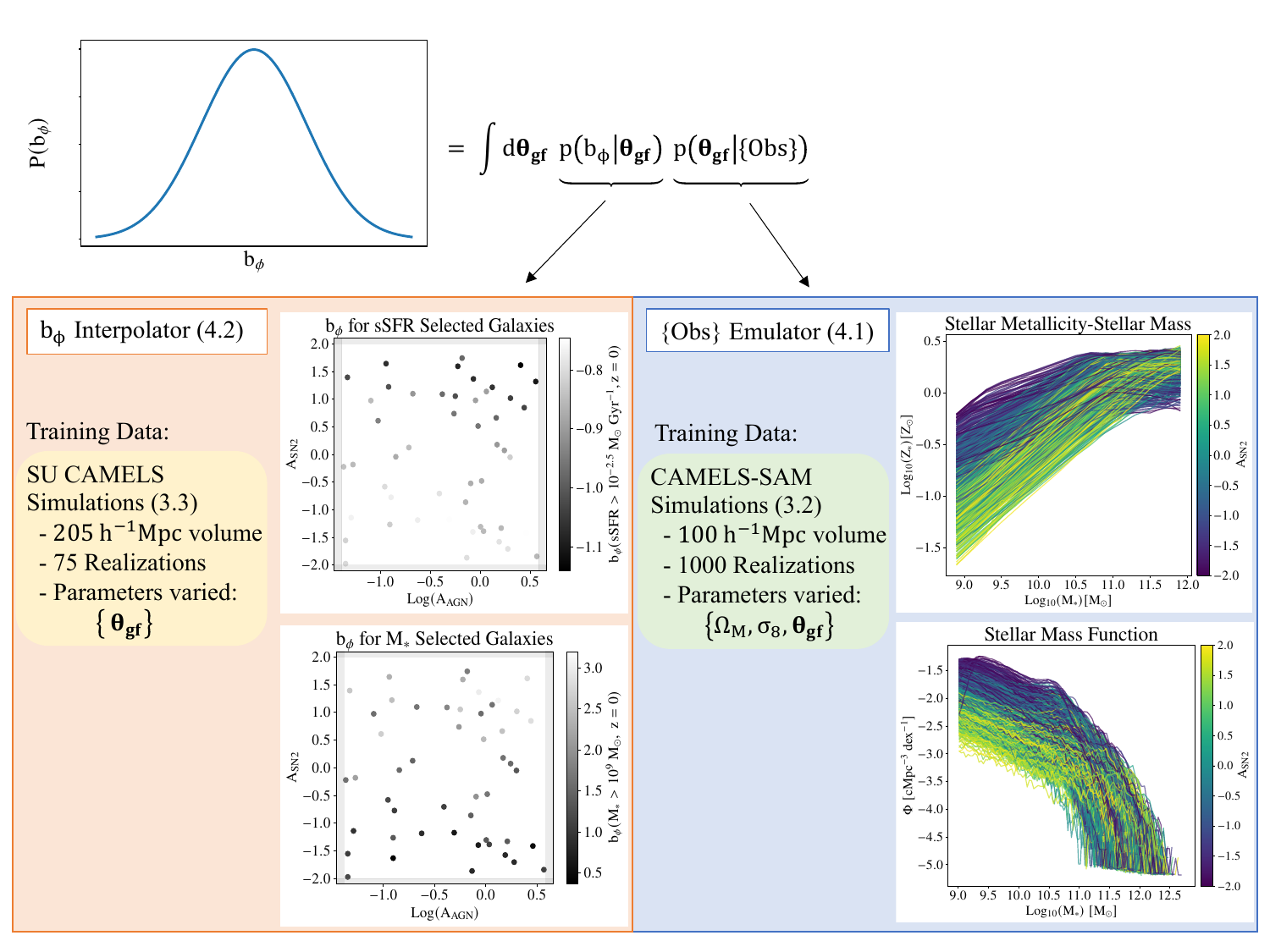} 
    \caption{Overview of the methodology used to construct the $b_{\phi}$ prior for some galaxy selection $S$ by marginalizing over galaxy formation uncertainties. \textit{Left:} The $b_{\phi}$ interpolator (Sec.~\ref{sec:Interp}) is trained on the 50 separate universe (SU) CAMELS-SAM simulations described in Sec.~\ref{sec:SUSam}. Three galaxy formation parameters ($\boldsymbol{\theta}_{\rm{gf}}$) are varied across these simulations and a value of $b_{\phi}$ is measured from each using the separate universe formalism (Sec.~\ref{sec:SU}) given some galaxy selection, such as stellar mass or specific star formation rate. \textit{Right:} A set of Gaussian Process Emulators (Sec.~\ref{sec:GPE}) is trained to predict galaxy formation observables using the CAMELS-SAM Latin Hypercube simulations of \cite{perez2023constrainingcosmologymachinelearning} (Sec.~\ref{sec:fullCAMELS}). Two cosmological parameters ($\Omega_{\rm{M}}$, $\sigma_8$) and three galaxy formation parameters ($\boldsymbol{\theta}_{\rm{gf}}$) are varied across 1000 realizations. Representative observables --- the stellar mass function (lower right) and the stellar metallicity-stellar mass relationship (upper right) --- illustrate the variation across the suite. The emulator enables MCMC sampling of $p\left(\boldsymbol{\theta}_{\rm{gf}}|\{Obs\}\right)$, and the interpolator maps sampled $\boldsymbol{\theta}_{\rm{gf}}$ values to $b_{\phi}$, yielding the marginalized prior $p\left(b_{\phi}|Obs\right)$ defined in eq.~\ref{eqn1:bphiPrior}.} \label{fig:networkDiagram} 
\end{figure}

Equation~\ref{eqn1:bphiPrior} is evaluated by combining two tools, each trained on a complementary simulation suite. A Gaussian Process Emulator, trained on galaxy formation observables, such as the stellar mass function and stellar metallicity-stellar mass relationship, from the CAMELS-SAM suite~\cite{perez2023constrainingcosmologymachinelearning}, enables MCMC sampling of $p\left(\boldsymbol{\theta}_{\rm{gf}}|\{Obs\}\right)$ at arbitrary points in the galaxy formation parameter space. A radial basis function interpolator, trained on the separate universe simulations of~\cite{perez2026impactgalaxyformationgalaxy}, defines a smoothed, deterministic mapping from $\boldsymbol{\theta}_{\mathrm{gf}}$ to $b_{\phi}$ and maps each posterior sample to a $b_{\phi}$ value. Since the parameter sets of~\cite{perez2026impactgalaxyformationgalaxy} constitute a subset of the CAMELS-SAM Latin hypercube, the prior volume is no longer regularly sampled, motivating the emulator-based MCMC approach. The resulting distribution constitutes the observation-conditioned $b_{\phi}$ prior $ p\left(b_{\phi}|\{Obs\}\right) $. The full methodology is summarized in Fig.~\ref{fig:networkDiagram}.   

The paper is organized as follows. Section~\ref{sec:SU} describes the separate universe formalism used to measure $b_{\phi}$ from simulations~\cite{Lazeyras_2016,Barreira_2020}. Section~\ref{sec:SimsDescription} describes the CAMELS-SAM simulations and their implementation of the separate universe approach. Section~\ref{sec:IntroML} details the design of the emulator and interpolator. Results  are presented in Section~\ref{sec:Results}, and Section~\ref{sec:Conclusion} discusses implications for future analyses.   

\section{Galaxy Bias from Separate Universe Simulations}\label{sec:SU}
The separate universe (SU) formalism provides a direct numerical approach to computing the response of galaxy number counts to long-wavelength primordial perturbations. It follows from the peak-background split~\cite{1984ApJ...284L...9K,1986ApJ...304...15B}, which decomposes a field into long- and short-wavelength modes. On scales small relative to the long-wavelength modes, the latter acts as a local modification of the background cosmology. First applied to local PNG by~\cite{Dai_2015}, the SU approach exploits the squeezed-limit form of the primordial bispectrum, in which a large-scale mode modulates the power spectrum of two short-scale modes. This results in a position-dependent modification of the primordial scalar power spectrum amplitude $\mathcal{A}_s$. The locally modified power spectrum at position $\boldsymbol{x}$ takes the form~\cite{Desjacques_2018} 
\begin{equation} 
    P_{\phi \phi} \left( k_{\rm{short}},z | \boldsymbol{x}\right) = P_{\phi \phi}\left(k_{\rm{short}},z\right)\lbrack 1 + 4f^{\rm{loc}}_{\rm{NL}}\phi\left(\boldsymbol{x}\right) \rbrack. 
\end{equation}  
Galaxies embedded within these long-wavelength perturbations evolve as if in a separate cosmology with a rescaled primordial power spectrum amplitude~\cite{PhysRevD.77.123514,Slosar_2008}:
\begin{equation}\label{eq:rescaledPrimordialAmp} 
    \tilde{\mathcal{A}_s} = \mathcal{A}_s \lbrack 1 + \underbrace{4f^{\rm{loc}}_{\rm{NL}}\phi_L}_ {\delta\mathcal{A}_s}\rbrack
\end{equation} 
where $\delta \mathcal{A}_s$ denotes the fractional modification to $\mathcal{A}_s$ induced by the long-wavelength PNG mode.

The bias parameter $b_{\phi}$, defined as $b_{\phi} \equiv \frac{d\ln n_g}{d\left(f^{\rm{loc}}_{\rm{NL}}\phi_L\right)}$, quantifies the response of galaxy number counts to long-wavelength primordial perturbations. Since galaxies within these perturbations form in an effectively rescaled cosmology, $b_{\phi}$ reflects the sensitivity of galaxy formation to the modified short-scale power spectrum amplitude $\delta\mathcal{A}_s$. Equivalently,  at redshift $z$,
\begin{equation}\label{biasDefAs} 
    b_{\phi} \left(z\right) = 4 \frac{\rm{d} \ln n_g\left(z\right)}{\rm{d} \left(\delta \mathcal{A}s\right)} \Bigg|_{\delta \mathcal{A}_s = 0}\ , 
\end{equation}  
where the factor of 4 arises from the relation $\delta\mathcal{A}_{s} = 4f_{\rm{NL}}^{\rm{loc}}\phi_{L}$ in eq.~\ref{eq:rescaledPrimordialAmp}.

For a galaxy selection $\mathrm{S}$, we evaluate eq.~\ref{biasDefAs} as a finite difference of galaxy number counts $N_g$ from simulations with bracketing values of $\mathcal{A}_s$:
\begin{equation} 
    b_{\phi}^{\rm{high,low}} \left(z,\mathrm{S}\right) = \frac{4}{\delta \mathcal{A}_s^{\rm{high,low}}} \left[ \frac{N_g^{\rm{high,low}}\left(z,\mathrm{S}\right)}{N_g^{\rm{fiducial}}\left(z,\mathrm{S}\right)} - 1\right]
\end{equation}  
where superscripts denote simulations run with $\mathcal{A}_s$ above and below the fiducial value respectively. We average the two estimates to obtain  
\begin{equation}
    b_{\phi}\left(z, \mathrm{S}\right) = \frac{b_{\phi}^{\rm{high}}\left(z,\mathrm{S}\right)+b_{\phi}^{\rm{low}}\left(z,\mathrm{S}\right)}{2}\ . 
\end{equation} 
Following~\cite{Barreira_2020, perez2026impactgalaxyformationgalaxy} we estimate the associated uncertainty as 
\begin{equation}\label{eq:measErr}
    \sigma_{b_{\phi}}\left(z,\mathrm{S}\right) = \frac{|b_{\phi}^{\rm{high}}\left(z,\mathrm{S}\right) - b_{\phi}^{\rm{low}}\left(z,\mathrm{S}\right)|}{2}\ .
\end{equation}
Barreira et al.~\cite{Barreira_2020} established that Poisson errors overestimate $\sigma_{b_{\phi}}$ by a factor of approximately 2, a consequence of the SU simulations sharing the same initial conditions; see their Sec. 2.2 for a detailed discussion.

\section{Simulations}\label{sec:SimsDescription}

This section provides an overview of the simulation suites used in this analysis. Section~\ref{sec:SAMs} describes the Santa Cruz semi-analytic model and its validation against observations and hydrodynamical simulations. Section~\ref{sec:fullCAMELS} presents the CAMELS-SAM suite and the parametrization of galaxy formation variations therein. Section~\ref{sec:SUSam} describes the separate universe simulations used to measure $b_{\phi}$.

\subsection{Semi-Analytic Models of Galaxy Formation}\label{sec:SAMs}  

Semi-analytic models (SAMs) are a well-established framework for incorporating galaxy formation physics into cosmological simulations~\cite{1991ApJ...379...52W,10.1093/mnras/264.1.201,10.1093/mnras/271.4.781,1999MNRAS.310.1087S}. Within a SAM, dark matter halo assembly is tracked through merger trees, constructed either from N-body simulations or from semi-analytic prescriptions based on the Press-Schechter formalism~\cite{10.1093/mnras/262.3.627,1993MNRAS.264..201K, 10.1093/mnras/271.3.676, 10.1046/j.1365-8711.1999.02154.x,2008MNRAS.383..557P}.  

Built upon merger trees, a SAM operates as a reservoir-flow framework, solving a set of coupled ordinary differential equations governing the exchange of baryonic material between the intergalactic medium (IGM), circumgalactic medium (CGM), interstellar medium (ISM), and stars. Gas cooling from the CGM into the ISM provides the fuel for star formation, which is modeled as a function of the cold gas surface density following a Schmidt-Kennicutt-type relation. Stellar evolution returns mass and metals into the ISM and CGM, while massive stars drive galactic winds parameterized by eq.~\ref{eq:StellarFeedback}, and $\dot{m}_*$ is set by the star formation prescription. Supermassive black holes grow via both cold gas accretion during mergers and the jet-mode accretion described by eq.~\ref{eq:AGNFeedback}, with $M_{\mathrm{BH}}$ evolving self-consistently throughout the simulation. The key baryonic outputs of the model --- stellar mass, star formation rate, metallicity, and black hole mass --- are therefore jointly determined by the interplay of these processes. While different SAMs share broad assumptions --- such as a common cooling and accretion prescription for CGM gas --- they differ in their treatment of star formation, stellar feedback, black hole growth, and AGN feedback. Despite these differences, multiple comparisons have demonstrated that diverse SAM implementations reproduce key observational quantities with broadly consistent results~\cite{Lu_2014,annurev:/content/journals/10.1146/annurev-astro-082812-140951,10.1093/mnras/stx3274}, and that SAM predictions are in reasonable agreement with those from hydrodynamical simulations~\cite{annurev:/content/journals/10.1146/annurev-astro-082812-140951}.  

The computational efficiency of SAMs relative to hydrodynamical simulations has enabled systematic exploration of large simulation volumes and broad galaxy formation parameter spaces. The CAMELS-SAM suite~\cite{shao2022robustfieldlevelinferencedark}, a component of the CAMELS project~\cite{Villaescusa_Navarro_2021,Villaescusa_Navarro_2023}, employs the Santa Cruz (SC) SAM~\cite{1999MNRAS.310.1087S, 10.1111/j.1365-2966.2008.13805.x, 10.1093/mnras/stv1877,10.1093/mnras/stab231}. Gabrielpillai et al.~\cite{10.1093/mnras/stac2297} demonstrated that this implementation reproduces key observational scaling relations --- including the stellar mass function, stellar-to-halo mass relationship, stellar metallicity-stellar mass relationship, bulge mass-black hole mass relationship, and cold gas-stellar mass relationship --- in good agreement with IllustrisTNG predictions~\cite{10.1093/mnras/stac2297}.

The CAMELS-SAM simulations were designed to probe the sensitivity of galaxy formation to variations in stellar and AGN feedback. Stellar feedback parametrizes two regimes of baryon cycling inefficiency: reduced star formation efficiency within the CGM, and suppressed stellar and baryon fractions in galactic subhalos relative to the cosmic mean. In the SC-SAM, the mass outflow rate due to stellar feedback scales with the depth of the gravitational potential well: 
\begin{equation}\label{eq:StellarFeedback} 
    \dot{m}_{\rm{out}} = \left(\epsilon_{\rm{SN}}\right)\left(\frac{V_0}{V_c}\right)^{\left(\alpha_{\rm{rh}}\right)}\dot{m_*} 
\end{equation}
where $\epsilon_{\rm{SN}}$ and $\alpha_{\rm{rh}}$ govern the normalization and slope of the stellar feedback scaling, $\\ V_0~=~200\,\text{km/s}$ is the threshold velocity for gas ejection from a halo, $V_c$ is the halo circular velocity, and $\dot{m}_*$ is the instantaneous star formation rate.  

AGN feedback in the SC-SAM describes the coupling of energy from radiatively inefficient accretion onto black holes to the hot halo gas through radio jet heating, regulating the growth of black holes. The jet-mode accretion rate is modeled as 
\begin{equation}\label{eq:AGNFeedback} 
    \dot{m}_{\rm{radio}} = \left(\kappa_{\rm{radio}}\right) \left[\frac{kT}{\Lambda\lbrack T,Z_h\rbrack}\right]\left(\frac{M_{\rm{BH}}}{10^8 M_{\odot}}\right) 
\end{equation}  
where $\kappa_{\rm{radio}}$ controls the jet-mode feedback efficiency, $T$ is the gas temperature at the Bondi radius, and $\Lambda\lbrack T,Z_h\rbrack$ is the temperature- and metallicity-dependent cooling function.   

The fiducial SC-SAM parameters $\epsilon_{\rm{SN}} = 1.7$, $\alpha_{\rm{rh}} = 3.0$, and $\kappa_{\rm{radio}} = 0.002$ were calibrated through iterative comparison with observational benchmarks~\cite{10.1111/j.1365-2966.2008.13805.x,10.1093/mnras/stv1877,10.1093/mnras/stab231,10.1093/mnras/sty3241}, including the stellar mass function, cold gas fraction, stellar mass-metallicity relation, and the black hole mass-bulge mass relation~\cite{10.1093/mnras/stt1607, Moustakas_2013, 10.1111/j.1365-2966.2012.20340.x, 10.1093/mnras/stx1172, 10.1093/mnras/sty089, calette2018hih2tostellarmasscorrelations, 10.1111/j.1365-2966.2005.09321.x, Kirby_2011, Schutte_2019,McConnell_2013, Kormendy_2013}. 

\subsection{CAMELS-SAM Simulation Suite}\label{sec:fullCAMELS}  
The CAMELS-SAM suite~\cite{Villaescusa_Navarro_2021,Villaescusa_Navarro_2023}, extends the CAMELS project by incorporating the SC-SAM to probe galaxy formation physics across a wide parameter space at substantially reduced computational cost compared to hydrodynamical simulations. Perez et al.~\cite{perez2023constrainingcosmologymachinelearning} used this suite to demonstrate the power of galaxy clustering statistics for jointly constraining cosmological and astrophysical parameters with neural networks.

The Latin Hypercube (LH) suite consists of 1005 N-body simulations of volume $\\ \left(100\ h^{-1}\, \text{cMpc}\right)^3$ with $N=640^3$ dark matter particles, compared to the $\left(25\ h^{-1}\,\text{cMpc}\right)^3$ volume with $N=256^3$ particles of the original CAMELS boxes. The N-body simulations were run with AREPO using an IllustrisTNG-like configuration, adopting a flat $\Lambda$CDM cosmology with fixed secondary parameters $\Omega_{\rm{b}} = 0.049$, $h = 0.6711$, $n_s = 0.9624$, $\sum m_{\nu} = 0.0\ \text{eV}$, and $w = -1$, while varying $\Omega_{\rm{M}} \in \left[ 0.1, 0.5\right]$ and $\sigma_8 \in \left[0.6,1.0\right]$. Halos and merger trees were identified using ROCKSTAR~\cite{Behroozi_2013_rockstar} and CONSISTENTTREES~\cite{Behroozi_2013_consistent}.  

Three dimensionless parameters --- $\rm{A_{SN1}}$, $\rm{A_{SN2}}$, and $\rm{A_{AGN}}$ --- were introduced as multiplicative pre-factors to the SC-SAM feedback equations to explore the impact of astrophysical uncertainties on some of the most important feedback processes in galaxy formation. The modified stellar feedback equation becomes
\begin{equation}\label{eq:stellarFeedbackCamelsSAM}
    \dot{m}_{\rm{out}} = \left(\epsilon_{\rm{SN}} \times \rm{A_{SN1}}\right)\left(\frac{V_0}{V_c}\right)^{\left(\alpha_{\rm{rh}}+\rm{A_{SN2}}\right)}\dot{m_*}\ . 
\end{equation}  
and the AGN jet-mode accretion rate becomes 
\begin{equation}\label{eq:AGNFeedbackCamelsSAM}
    \dot{m}_{\rm{radio}} = \left(\kappa_{\rm{radio}}\times\rm{A_{AGN}}\right) \left[\frac{kT}{\Lambda\left[ T,Z_h\right]}\right]\left(\frac{M_{\rm{BH}}}{10^8 M_{\odot}}\right)\ . 
\end{equation}
Variation in $\rm{A_{ASN1}}$ and $\rm{A_{SN2}}$ will modify the mass outflow rate of galaxies due to supernovae radiation from massive stars within the SC-SAM, and variation in $\rm{A_{AGN}}$ will modify the strength of feedback from the jet mode within the SC-SAM. 

The parameters $\rm{A_{AGN}}$ and $\rm{A_{SN1}}$ were sampled log-uniformly over $\left[ 0.25, 4.0 \right]$ while $\rm{A_{SN2}}$ was sampled linearly over $\left[ -2.0, 2.0 \right]$. Prior ranges and fiducial values for all varied parameters are summarized in Table~\ref{tab:VariedParamsSAM}.

\begin{table}[h] 
\centering 
\begin{tabular}{|c|c|c|} 
    \hline Parameter& Prior Range & Fiducial Value \\ 
    \hline 
    $\Omega_{\rm{M}}$ & $\mathcal{U}\lbrack 0.1, 0.5\rbrack$ & $0.3$\\ 
    $\sigma_8$ & $\mathcal{U}\lbrack 0.6, 1.0\rbrack$ & $0.8$\\ 
    $\rm{A_{AGN}}$ & $\mathcal{U}\lbrack \rm{log}\left(0.25\right), \rm{log}\left(4.0\right)\rbrack$ & $\mathrm{log}\left(1.0\right)$ \\ 
    $\rm{A_{SN1}}$ & $\mathcal{U}\lbrack \rm{log}\left(0.25\right), \rm{log}\left(4.0\right)\rbrack$ & $\mathrm{log}\left(1.0\right)$\\ 
    $\rm{A_{SN2}}$ & $\mathcal{U}\lbrack -2.0, 2.0\rbrack$ & $0.0$\\ 
    \hline 
\end{tabular} 
\caption{Prior ranges and fiducial values of the cosmological and galaxy formation parameters varied in the CAMELS-SAM Latin Hypercube suite. Parameters $\rm{A_{AGN}}$ and $\rm{A_{SN1}}$ are sampled in uniformly log space; all others are sampled linearly.}\label{tab:VariedParamsSAM}
\end{table} 

\subsection{Separate Universe Simulations to Measure \texorpdfstring{$b_{\phi}$}{bphi}}\label{sec:SUSam}  

To measure $b_{\phi}$ across a range galaxy formation realizations, Perez et al.~\cite{perez2026impactgalaxyformationgalaxy} constructed an additional suite of separate universe (SU) simulations at larger volume, designed to capture the large-scale response necessary for accurate $b_{\phi}$ measurements. These simulations employ the same SC-SAM implementation as the CAMELS-SAM suite (Sec.~\ref{sec:fullCAMELS}) and were run on larger N-body boxes to exploit the SU formalism described in Sec.~\ref{sec:SU}. The cosmological parameters are fixed at their fiducial values ($\Omega_{\rm{M}}=0.3$,\ $\sigma_8=0.8$). 

The SU suite comprises three N-body simulations of volume $\left(205\ h^{-1}\, \text{Mpc} \right)^3$ box with $N = 1280^3$ dark matter particles, sharing the same Gaussian initial conditions but run with $\sigma_8 \in \{0.72,0.80,0.88\}$ corresponding to the low, fiducial, and high amplitude realizations respectively.   
\begin{figure}[h] 
\centering 
    \includegraphics[trim={2cm 2cm 2.5cm 4cm},clip,width=1.\textwidth,height=12cm]{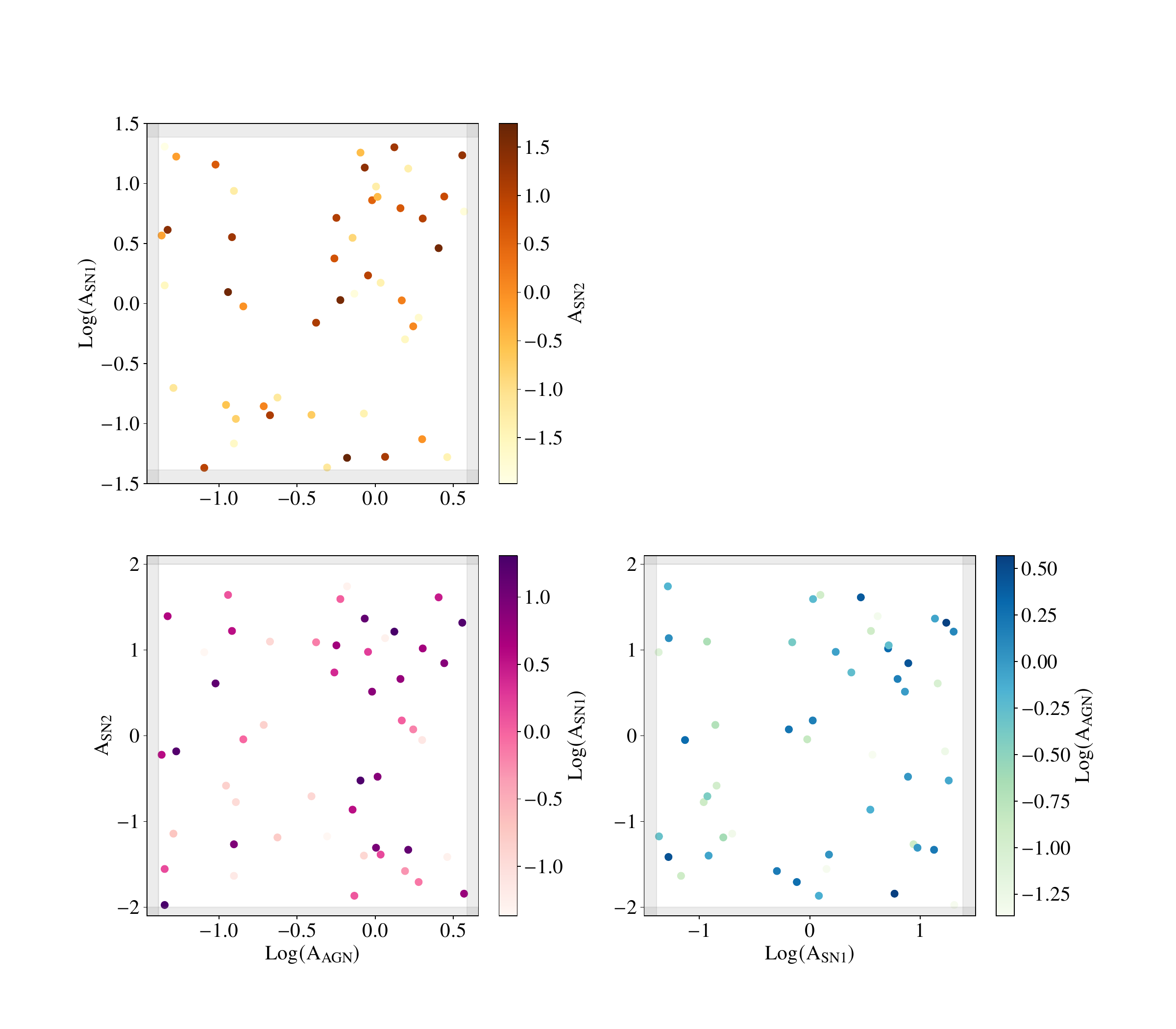} 
    \caption{Distribution of galaxy formation parameters for the 50 SU CAMELS-SAM simulations retained after the $\mathrm{A_{AGN}}>1.8$ cut, confirming coverage of the full CAMELS-SAM prior range (prior bounds indicated by the gray shaded region). Parameters $\rm{A_{AGN}}$ and $\rm{A_{SN1}}$ are shown in log space consistent with their log-uniform sampling.}\label{fig:paramDistScatter} 
\end{figure}  
The SC-SAM was applied to each of the three N-body boxes for 75 sets of galaxy formation parameters $\{\rm{A_{AGN}},\  \rm{A_{SN1}},\ \rm{A_{SN2}}\}$ chosen as a subset of the CAMELS-SAM LH parameters spanning the full prior range of Table~\ref{tab:VariedParamsSAM}. For each simulation, Perez et al.~\cite{perez2026impactgalaxyformationgalaxy} measured a suite of galaxy formation observables and quantified the realism of each realization by computing a normalized $\chi^2$ against observational data. Their Fig. 4 shows the resulting stellar mass function and stellar-to-halo mass relationship across the suite, shaded by realism. This realism weighting informs the $b_{\phi}$ analysis presented in this work. Following~\cite{perez2026impactgalaxyformationgalaxy}, we exclude simulations with $\rm{A_{AGN}}>1.8$, as this regime produces unphysical galaxy properties; we impose this cut consistently throughout all MCMC analyses in Sec.~\ref{sec:Results}, reducing the retained suite to 50 realizations. Figure~\ref{fig:paramDistScatter} confirms that these 50 parameter sets provide adequate coverage of the full prior volume. 

\section{Constructing \texorpdfstring{$\boldsymbol{b_{\phi}}$}{bphi} Priors Conditioned on 
Observables}\label{sec:IntroML}

The practical evaluation of $p(b_\phi |\{Obs\})$
(eq.~\ref{eqn1:bphiPrior}) requires measurements of observables and $b_{\phi}$ values across the full galaxy formation parameter space. Since the computational cost of the separate universe SAMs precludes dense sampling, $p(b_\phi|\theta_{\rm{gf}})$ is available only on an irregular subset of the CAMELS-SAM hypercube. To address this, we use a Gaussian Process Emulator trained on the CAMELS-SAM simulations (Sec.~\ref{sec:fullCAMELS}) to predict galaxy evolution observables at arbitrary parameter locations, enabling MCMC exploration of $p(\boldsymbol{\theta}_{\rm{gf}}|\{Obs\})$. We then use a radial basis function interpolator, trained on the SU simulations of Sec.~\ref{sec:SUSam}, to map each posterior sample to a $b_{\phi}$ prediction, yielding the observation-conditioned distribution $p(b_{\phi}|\{Obs\})$.

\subsection{Emulation of Galaxy Formation Observables}\label{sec:GPE}

During MCMC sampling, we vary the parameters $\rm{A_{AGN}},\ \rm{A_{SN1}},\ \text{and}\ \rm{A_{SN2}}$ within the prior ranges of Table~\ref{tab:VariedParamsSAM}. For each proposed parameter set proposed during sampling, we predict the corresponding galaxy observable using a Gaussian Process Emulator (GPE) implemented with scikit-learn's \texttt{GaussianProcessRegressor}~\cite{scikit-learn}. 

A GPE is specified by a mean function $m\left(x\right)$ --- set to zero throughout --- and a kernel function $k\left(x,x^{\prime}\right)$ encoding the covariance structure of the output:
\begin{equation} 
    F \left(x\right) \sim GP\left( m\left(x\right), k\left(x,x^{\prime}\right)\right) 
\end{equation}  
where the functional form of $k\left(x,x^{\prime}\right)$ will vary depending on the data set. After comparing the performance of available kernels in the scikit-learn library, we adopt a kernel combining a Radial Basis Function (RBF) and White Noise component:
\begin{equation} 
 k\left(x,x^{\prime}\right) = \text{exp}^{\lbrack \frac{-d \left(x,x^{\prime}\right)^2}{2 \ell^2}\rbrack} + \sigma^2 \delta^{\rm{D}}_{x,x^{\prime}} 
\end{equation}  
where $\ell$ is the characteristic length scale, $\sigma^2$ is the noise variance, $d$ is the Euclidean distance between input points, and $\delta^D$ is the Dirac delta function. We optimize both hyperparameters by maximizing the log marginal likelihood during training.  

We measure galaxy formation observables $\{Obs\}$ as a function of mass from each LH simulation (Sec.~\ref{sec:fullCAMELS}) to build the training set. We train independent GPEs to predict each observable at each mass bin as a function of $\{\Omega_{\rm{M}},\ \sigma_8,\ \rm{A_{AGN}},\ \rm{A_{SN1}},\ \rm{A_{SN2}}\}$. We standardize all input parameters using scikit-learn’s \texttt{StandardScaler} prior to training, transforming the inputs to zero mean and unit variance; GPEs are known to perform better on inputs following this distribution, as the transformation removes scaling artifacts. We transform predictions back to physical units before use. 

We partition the full simulation suite into a training set (85\%) and a held-out validation set (15\%). We assess GPE performance on the validation set by computing residuals between predicted and true observables, and propagate these residuals into the total covariance matrix as described in Sec.~\ref{subsec:FidErrEst}.  

\subsection{\texorpdfstring{$b_{\phi}$}{bphi} Interpolation}\label{sec:Interp} 

We construct a radial basis function interpolator to predict $b_{\phi}$ from the galaxy formation parameters $\{\rm{A_{AGN}},\  \rm{A_{SN1}},\ \rm{A_{SN2}}\}$, using SciPy's \texttt{RadialBasisFunctionInterpolator}~\cite{2020SciPy-NMeth} for its flexibility with high-dimensional, non-uniformly sampled data. The training set consists of the 50 galaxy formation parameter sets input to the SU simulations (Sec.~\ref{sec:SUSam}) and the corresponding $b_{\phi}$ value measured from each using the formalism of Sec.~\ref{sec:SU}. 

We define the galaxy sample in each simulation to replicate the selection criteria of the DESI ELG sample as characterized in \cite{Yuan_2022}, with stellar mass $10^{9.5}\ M_{\odot} < M_* < 10^{11.}\ M_{\odot}$ and specific star formation rate $10^{-3.5}\ \rm{Gyr}^{-1} < sSFR < 10^{1.5}\ \rm{Gyr}^{-1}$. This yields a single $b_{\phi}$ measurement per simulation, in contrast to the binned approach of~\cite{perez2026impactgalaxyformationgalaxy} which produces multiple measurements per realization.

\section{Results}\label{sec:Results}

We use the trained emulators and interpolator described in Sections~\ref{sec:GPE} and~\ref{sec:Interp} to run MCMC analyses and derive $b_{\phi}$ posterior distributions conditioned on three galaxy formation observables. We use the public \texttt{emcee} sampler \cite{Foreman_Mackey_2013} to sample the posterior distribution $p\left(\{\rm{A_{AGN}}, \rm{A_{SN1}}, \rm{A_{SN2}}\}|\{Obs\}\right)$ conditioned on the stellar mass function (SMF), the stellar-to-halo mass relationship (SHMR), and the stellar metallicity-stellar mass relationship ($M_*-Z_*$). We select these observables for their sensitivity to variations in the galaxy formation parameters as established in~\cite{perez2023constrainingcosmologymachinelearning} (see their Appendix A).

The SMF --- the number density of galaxies as a function of stellar mass --- constrains the growth of stellar mass and the history of star formation. We train independent GPEs to predict the log normalized number density in each of 20 equally-spaced stellar mass bins spanning $10^{9}\ \text{M}_{\odot} < M_* < 10^{11}\ \text{M}_{\odot}$, matching the binning of Bernardi et al.~\cite{Bernardi_2017}. We exclude bins above $M_* > 10^{11}\ \text{M}_{\odot}$ as the inferred SMF in this regime is sensitive to assumptions in photometric stellar mass estimation~\cite{Bernardi_2017}. 

The SHMR\footnote{We note that the stellar-to-halo mass ratio is not a direct observable like the SMF or $M_*-Z_*$, as it is not measured directly from observations. Instead, it is the best-fit of a semi-empirical model for the implied galaxy-halo connection of the Universe given a variety of observables, such as the stellar mass function and specific star formation rates~\cite{10.1093/mnras/stx1172} from different data sets. For this reason, we treat it as a pseudo-observable in the remainder of this work.} quantifies the efficiency of stellar mass assembly as a function of halo mass and is sensitive to both stellar and AGN feedback across the relevant mass range . We train emulators to predict the log stellar-to-halo mass ratio in 41 equally-spaced halo mass bins spanning $10^{10.7}\ \text{M}_{\odot} < M_{\rm{halo}} < 10^{14.7}\ \text{M}_{\odot}$, motivated by the observational data of Rodr\'iguez-Puebla et al.~\cite{10.1093/mnras/stx1172}.

The stellar metallicity-stellar mass relationship quantifies the mean metal abundance of galaxies as a function of stellar mass, tracing metal retention across the galaxy population. We train a emulators to predict the log stellar metallicity in 16 equally-spaced stellar mass bins spanning $ 10^{9}\ \rm{M}_{\odot}<M_*<10^{12}\ \rm{M}_{\odot}$, following the measurement of Gallazzi et al.~\cite{10.1111/j.1365-2966.2005.09321.x}.

\subsection{Validation on Simulated Data}\label{sec:FiducialDataVec}
\subsubsection{Estimation of Simulated Data Vector and Covariance Matrix}\label{subsec:FidErrEst}
We first validate the methodology against the CAMELS-SAM simulations, by verifying recovery of the fiducial galaxy formation parameters before applying the framework to observed data. We construct a mock data vector $\widehat{Obs}_{i}$ by perturbing each observable measured from the fiducial simulation $Obs_{\rm{fid}}$ with Gaussian noise: 
\begin{equation}\label{eq:fidData}
    \widehat{Obs}_{i} = Obs_{\mathrm{fid},i} + \mathcal{N}\left(0,\sigma_{\mathrm{data},i}\right)\ .
\end{equation}
where $Obs_{\mathrm{fid},i}$ is the value of our fiducial observable in mass bin $i$, and $\sigma_{\mathrm{data},i}$ is the observational uncertainty in mass bin $i$ taken from the respective data reference: \cite{Bernardi_2017} for the SMF, \cite{10.1093/mnras/stx1172} for the SHMR, and \cite{10.1111/j.1365-2966.2005.09321.x} for $M_*-Z_*$.

We assume a Gaussian likelihood with a total diagonal covariance matrix that incorporates four contributions:
\begin{equation}\label{eq:totalCovFid}
    C_{\rm{tot}} = \mathrm{diag}\left(\sigma^2_{\mathrm{data},i}\right) + C_{\rm{GP}} + C_{\rm{fid}} + C_{\rm{sim}}
\end{equation}
representing observational measurement uncertainties, GPE emulation errors, fiducial simulation sampling variance, and CAMELS-SAM simulation variance (including finite training set noise) respectively. We estimate GPE errors from the residuals between emulator predictions and true observables on the held-out 15\% validation set; consistent with the per-mass-bin GPE training, we treat these contributions as uncorrelated across mass bins and include them in $C_{\rm{GP}}$ diagonally. We estimate the fiducial simulation sampling variance in mass bin $i$ as:
\begin{equation}\label{eq:variance}
\operatorname{Var}_{{\rm{fid}}, i}\left(x\right) = \frac{\sigma_{\rm{MAD},\textit{i}}^2}{N_{i}-1}
\end{equation}
where $N_{i}$ is the number of galaxies in each bin, $x$ represents a measured galaxy property in bin $i$, and $\sigma_{\rm{MAD},\textit{i}}^2$ is the Median Absolute Deviation (MAD) of that property in each bin, preferred over standard deviation for its robustness to outliers. 

The simulation variance term $C_{\rm{sim}}$ captures both the LH simulation sampling variance and the finite training set contribution. We extend eq.~\ref{eq:variance} to compute the per-bin variance for each simulation $K$ as:
\begin{equation}\label{eq:varianceMultSim}
\operatorname{Var}_{\mathrm{sim},i,K}\left(x\right) = \frac{\sigma_{\mathrm{MAD},i,K}^2}{N_{i,K}-1}
\end{equation}
where $N_{i,K}$ is the number of galaxies per mass bin $i$ per simulation $K$ and $\sigma_{\mathrm{MAD},i,K}$ is the Median Absolute Deviation measured per mass bin per simulation $K$. We pool these across the suite to measure the variance in one bin $i$.

We estimate the finite training set contribution to $C_{\rm{sim}}$ from the squared residual between the fiducial observable and its nearest neighbor in the LH parameter space and add this diagonally.

\subsubsection{Constraints on \texorpdfstring{$b_{\phi}$}{bphi} For CAMELS-SAM}\label{subsec:fidConst}

We run three MCMC analyses, each conditioning on one of the galaxy observables described above, using the \texttt{emcee} sampler~\cite{Foreman_Mackey_2013}. We fix the cosmological parameters at their fiducial values ($\Omega_M = 0.3$,\ $\sigma_8 = 0.8$), vary $\rm{A_{AGN}}$ within $0.25<\rm{A_{AGN}}<1.8$ following \cite{perez2026impactgalaxyformationgalaxy}, and vary $\{\rm{A_{SN1}}, \rm{A_{SN2}}\}$ within the ranges of Table~\ref{tab:VariedParamsSAM}. We predict a $b_{\phi}$ value for each posterior sample via the interpolator of Sec.~\ref{sec:Interp}, and evaluate the resulting $b_{\phi}$ distributions for the ELG-like galaxy selection.

\begin{figure}[h]
    \centering
    \includegraphics[height=10cm, width=0.8\textwidth]{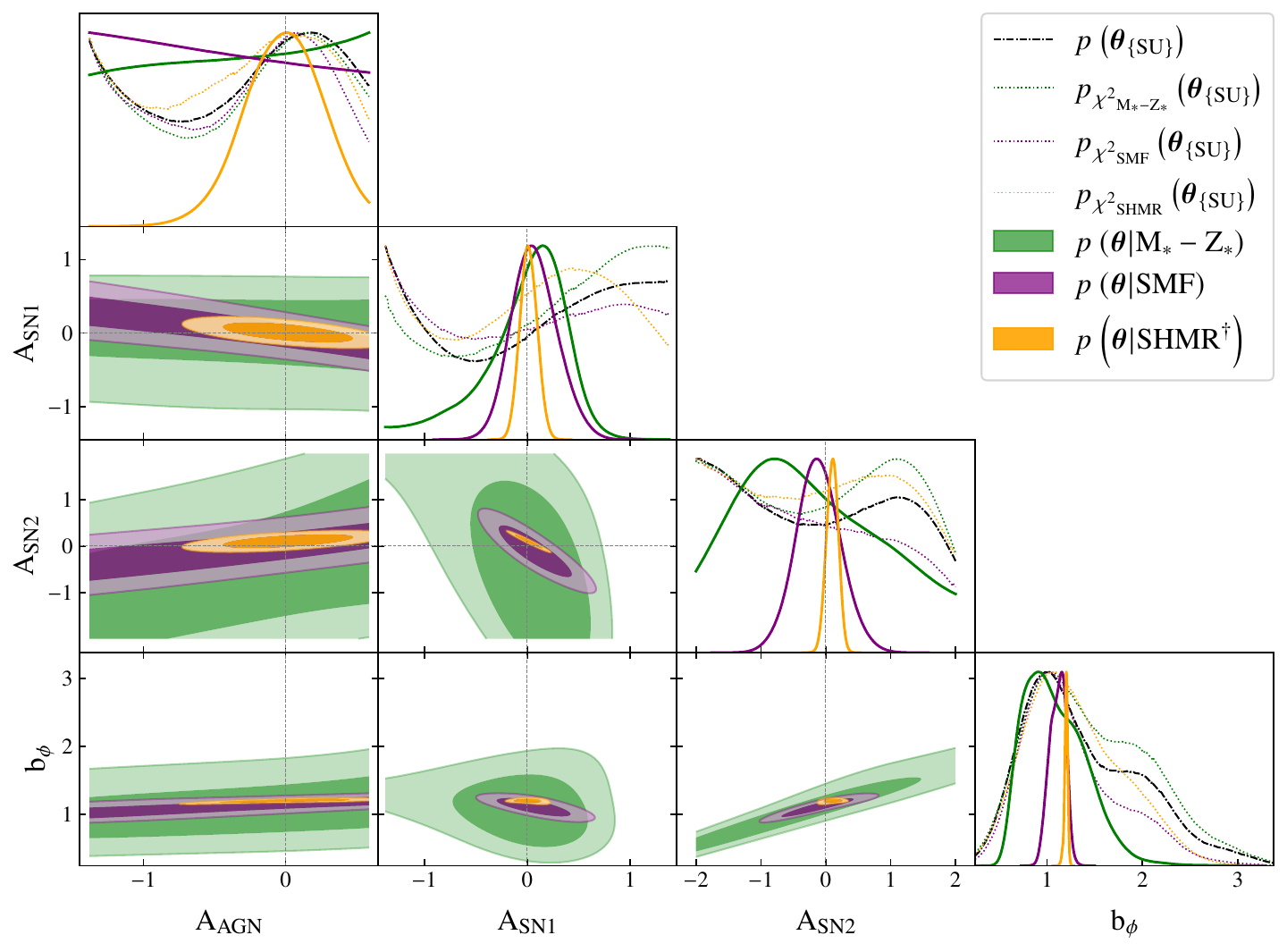}
    \caption{$b_{\phi}$ prior distributions conditioned on observables measured from the fiducial CAMELS-SAM simulation (Sec.~\ref{subsec:FidErrEst}). Solid contours show the galaxy formation parameter posteriors conditioned on: the stellar metallicity-stellar mass relationship (green, $M_*-Z_*$), the stellar mass function (purple, SMF), and the stellar-to-halo mass relationship (yellow, SHMR). Colored dashed lines show $\chi^2$-weighted parameter distributions from the 50 SU simulations; black dashed line shows the unweighted SU parameter distribution. Parameters $\rm{A_{AGN}}$ and $\rm{A_{SN1}}$ are shown in log space. Conditioning on galaxy formation observables yields narrower $b_{\phi}$ posteriors relative to the $\chi^2$-weighted distributions, with the largest reductions from the SMF and SHMR. See Table~\ref{tab:fidAnalysisSummary} for quantitative comparisons.}\label{fig:CornerPlotFidDataVec}
\end{figure} 

Figure~\ref{fig:CornerPlotFidDataVec} shows the galaxy formation parameter posteriors and the resulting $b_{\phi}$ distributions for the DESI-like galaxy sample defined in Sec.~\ref{sec:Interp}. The dashed lines show the prior parameter distribution weighted by the realism of each SU simulation $K$, quantified as:
\begin{equation}\label{eq:chi2}
    \chi^2_{K} = \left(Obs_{\rm{fid}} - Obs_{\mathrm{SU}_K}\right)C^{-1}\left(Obs_{\rm{fid}} - Obs_{\mathrm{SU}_K}\right)^T\ ,
\end{equation}
where the covariance used to compute the $\chi^2$ value for each simulation is 
\begin{equation}\label{eq:covarianceChi2}
    C = \mathrm{diag}\left(\sigma^2_{\mathrm{data},i}\right) + C_{\rm{fid}} + C_{\rm{sim}}
\end{equation}
and $Obs_{\rm{fid}}$ is the data vector measured from the fiducial simulation. We omit the GPE emulation error $C_{\rm{GP}}$ since the SU observables are measured directly from simulations rather than through the emulator. This $\chi^2$ formulation differs slightly from that used in~\cite{perez2026impactgalaxyformationgalaxy} in that we include simulation measurement variance and observational errors simultaneously, without averaging the $\chi^2$ value. We show the unweighted SU distribution as an additional comparison (black dashed line). 

The results are consistent with the sensitivity analysis of~\cite{perez2023constrainingcosmologymachinelearning}: we recover the fiducial value of $\rm{A_{SN1}}$ when conditioning on the SMF and SHMR, with a weaker but consistent constraint from $M_*-Z_*$. We find similar behavior for $\rm{A_{SN2}}$, as the SMF and SHMR are sensitive to this parameter across the relevant mass ranges; $M_*-Z_*$ provides weaker constraints due to large observational uncertainties at these scales.

Constraining power on the AGN feedback parameter $\rm{A_{AGN}}$ is limited when conditioning on the SMF and $M_*-Z_*$. In the case of the SMF, this follows from our exclusion of mass bins above $M_*> 10^{11}M_{\odot}$ --- the regime where $\rm{A_{AGN}}$ influences the stellar mass function amplitude, as established by~\cite{perez2023constrainingcosmologymachinelearning}. The SHMR retains sensitivity to $\rm{A_{AGN}}$ at halo masses $M_{\rm{halo}}> 10^{12}\ M_{\odot}$, where variations in AGN feedback shift the stellar-to-halo mass turnover point and modify the amplitude of the relationship at high masses ($M_{\rm{halo}}\sim10^{13}\ M_{\odot}$). The resulting tighter posteriors on the feedback parameters reduce the range of $b_{\phi}$ values consistent with observations, with the SHMR yielding the largest improvement owing to its additional sensitivity to $\rm{A_{AGN}}$.

Table~\ref{tab:fidAnalysisSummary} quantifies these results. The $\chi^2$-weighted distributions yield modest reductions in $\sigma_{b_{\phi}}$, with the largest improvements of $\Delta\sigma_{b_{\phi}}/\sigma_{b_{\phi,\mathrm{ref}}} = 28.7\%\ \text{and}\ 28.9\%$ from the SMF and SHMR respectively. Conditioning on observables via the MCMC framework yields reductions of $88.4\%$ and $96.7\%$ from the SMF and SHMR respectively, and $52.6\%$ from $M_*-Z_*$; the smaller improvement from $M_*-Z_*$ is consistent with the larger observational uncertainties in the Gallazzi et al.~\cite{10.1111/j.1365-2966.2005.09321.x} data at the relevant stellar mass scales.

\begin{table*}[h]
    \centering
    \begin{tabular*}{ \linewidth}{@{\extracolsep{\fill}} c|c|ccc|ccc}
        \multicolumn{8}{c}{\rule{0pt}{1cm}{\text{$b_{\phi}$ Priors Conditioned on Simulated Data}}\rule[-1ex]{0pt}{1cm}}  \\
        \hline
         \rule{0pt}{0.7cm}Metric & $\langle b_{\phi}\rangle_{\rm{SU}}$ & \multicolumn{3}{c|}{$\langle b_{\phi}\rangle_{{\chi^2}_{\{Obs\}}}$} & \multicolumn{3}{c}{$p\left(b_{\phi} |\{Obs\}\right)$} \\
         &  & SMF & SHMR & $M_*-Z_*$ & SMF & $\text{SHMR}^{\dagger}$ & $M_*-Z_*$\rule{0pt}{0.7cm} \\
         \hline \hline
         $\langle b_{\phi}\rangle $ & $1.43$ & $1.23$ & $1.28$ & $1.44$ & $1.11$ & $1.19$ & $1.09$ \\
         $\sigma_{b_{\phi}} $ & $0.69$ & $0.50$ & $0.49$ & $0.63$ & $0.08$ & $0.02$ & $0.33$ \\
         $\Delta\sigma_{b_{\phi}}/\sigma_{b_{\phi},\rm{ref}}$ & --- & $28.7\%$ & $28.9\%$ & $8.69\%$ & $88.4\%$ & $96.7\%$ & $52.6\%$\\
         K-L Divergence & --- & $0.12$ &$0.16$ & $0.05$ & $1.56$ & $2.89$ & $0.34$\\
         \hline
    \end{tabular*}
    \caption{Mean and symmetrized uncertainty of $b_{\phi}$ for different analyses. The first column reports the mean and standard deviation of the unweighted SU parameter distribution serving as our reference~\cite{perez2026impactgalaxyformationgalaxy}. Columns 2--4 give results for the $\chi^2$-weighted average over SU realizations; columns 5--7 give posteriors from the combined SU and CAMELS-SAM analyses.  
    Row 3 reports the fractional reduction in $\sigma_{b_{\phi}}$ relative to the unweighted distribution. Row 4 reports the Kullback-Leibler divergence $D_{\rm{KL}}\left(P\|Q\right)$ where $P$ is the (weighted) posterior and $Q$ the prior.\\
    $^\dagger$ $b_\phi$ inference conditioned on the observed SHMR is unreliable due to model-misspecification, resulting in poor goodness of fit.}\label{tab:fidAnalysisSummary}
\end{table*}
The Kullback-Leibler (K-L) divergence $D_{\rm{KL}}\left(P\|Q\right)$ provides an additional measure of information gain relative to the unweighted prior $Q$, where $P$ denotes the weighted or conditioned distribution. Values near zero indicate that $P$ is statistically indistinguishable from the unweighted prior, while larger values reflect information gain. The $\chi^2$-weighted distributions yield K-L values of $0.05$--$0.16$, confirming that simple simulation reweighting provides negligible constraining power beyond the prior. The MCMC-conditioned distributions yield K-L values of $1.56$ and $2.89$ for the SMF and SHMR respectively, indicating that the emulator-based MCMC approach extracts considerably more information from the galaxy formation observables.

All MCMC analyses prefer lower values of $b_{\phi}$ relative to the unweighted distribution, with normalized mean shifts of $-0.66\sigma_{\rm{sym}}$ to $-0.50\sigma_{\rm{sym}}$ for the SMF and SHMR.

\subsection{Application to Observed Data}\label{sec:ObsDataVec}

\subsubsection{Observed Data Vectors and Covariance Estimation}\label{subsec:DataErrEst}

We now apply the validated framework to observational data, replacing the simulated data vectors with the SMF from Bernardi et al.~\cite{Bernardi_2017}, the SHMR from Rodr\'iguez-Puebla et al.~\cite{10.1093/mnras/stx1172}, and the stellar metallicity-stellar mass relationship from Gallazzi et al.~\cite{10.1111/j.1365-2966.2005.09321.x}. These datasets were used in the original calibration of the SC-SAM fiducial parameters (Sec.~\ref{sec:SAMs}; see Appendix A of \cite{perez2023constrainingcosmologymachinelearning} for the complete list).

When applied to observational data, we simplify the covariance in  eq.~\ref{eq:totalCovFid} to:
\begin{equation}\label{eq:totalCovData}
    C_{\rm{tot}} = \mathrm{diag}\left(\sigma^2_{\mathrm{data},i}\right) + C_{\rm{GP}} + C_{\rm{sim}} 
\end{equation}
omitting $C_{\rm{fid}}$, the relevant uncertainty is the observational measurement error rather than the fiducial simulation sampling variance. We recompute the $\chi^2$ weights for the SU parameter distributions accordingly:
\begin{equation}\label{eq:chi2Data}
    \chi^2 = \left(Obs_{\rm{data}} - Obs_{\rm{SU}}\right)C^{-1}\left(Obs_{\rm{data}} - Obs_{\rm{SU}}\right)^T
\end{equation}
where $C_{\rm{fid}}$ has been removed from the covariance for the same reason as in eq.~\ref{eq:totalCovData}.

\subsubsection{Constraints on \texorpdfstring{$b_{\phi}$}{bphi} from Observed Data}

\begin{figure}[h]
    \centering
    \includegraphics[height=10cm, width=0.8\textwidth]{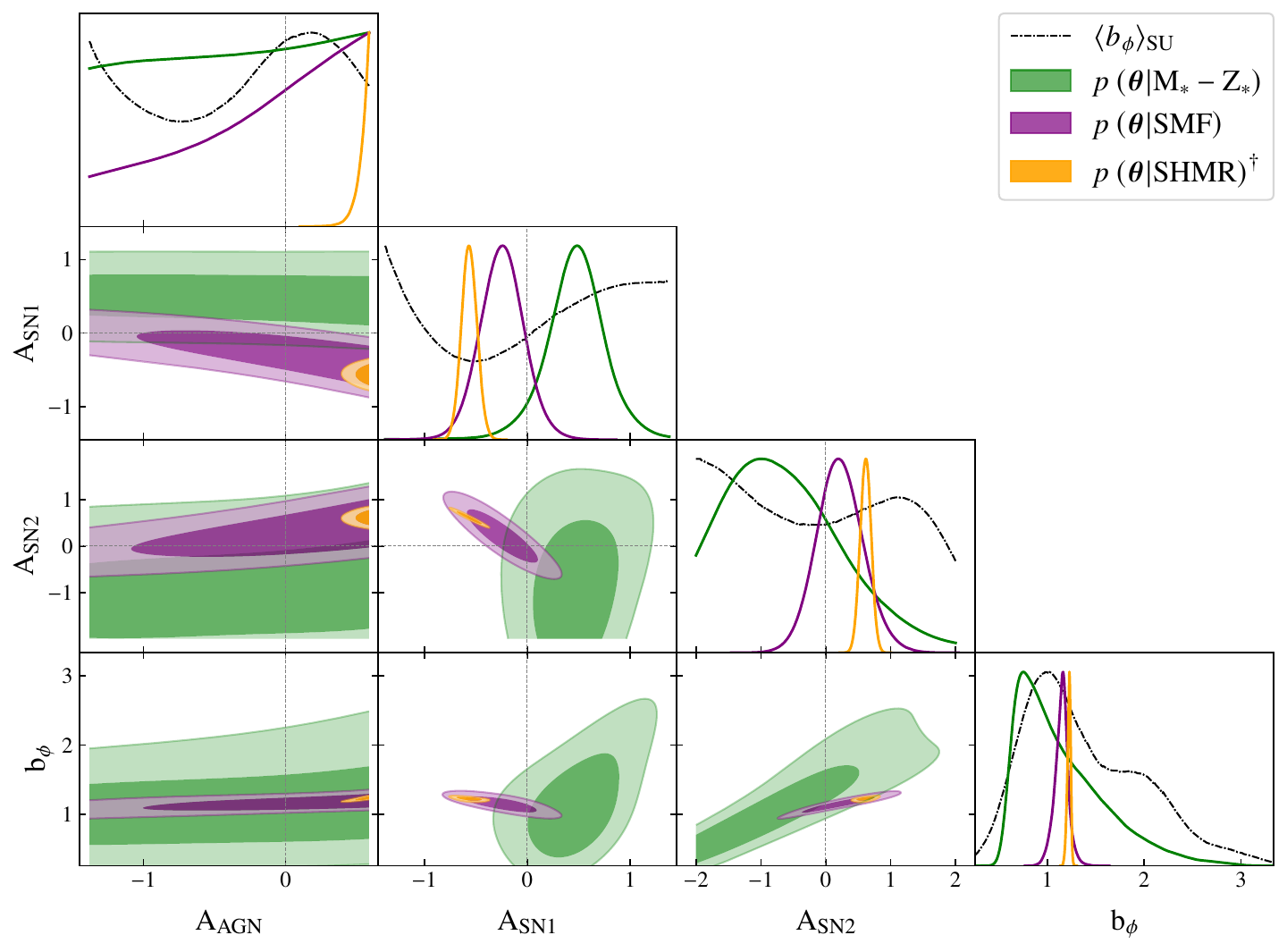}
    \caption{Same as Fig.~\ref{fig:CornerPlotFidDataVec}, but conditioned on observational data vectors (Sec.~\ref{subsec:DataErrEst}). The SHMR posterior (yellow) exhibits a marked shift in $\rm{A_{AGN}}$, driven by the known suppression of the high-mass region of the stellar-to-halo mass ratio in the fiducial SC-SAM relative to observations. Despite the shift in the galaxy formation posteriors, the resulting $b_{\phi}$ distributions remain mutually consistent across all three observables.\\
    $^\dagger$ $b_\phi$ inference conditioned on the observed SHMR is unreliable due to model-misspecification, resulting in poor goodness of fit.}\label{fig:CornerPlotObsDataVec}
\end{figure} 

We present the results in Fig.~\ref{fig:CornerPlotObsDataVec}. When conditioning on the SHMR, we find that the posterior mean of $\rm{A_{AGN}}$ shifts from its fiducial value of $\rm{A_{AGN}} = 0.98$ to $\rm{A_{AGN}} = 1.72$, approaching the prior boundary, accompanied by compensating shifts in $\rm{A_{SN1}}$ and $\rm{A_{SN2}}$. This shift in $\rm{A_{AGN}}$ reflects the known difficulty of reproducing the observed SHMR amplitude at high halo masses in the SC-SAM, where the SC-SAM struggles to sufficiently suppress the stellar-to-halo mass ratio at $M_{\rm{halo}}\gtrsim10^{13}M_{\odot}$ to match observations (see Fig. 4 in Appendix A of \cite{perez2023constrainingcosmologymachinelearning}). A similar behavior was observed in Pandya et al.~\cite{pandya2026introducingsapphirehybridphysicsinformed}, who studied the impact of applying systematic shifts to observable relations, measured from a different SAM with different physics, on astrophysical parameter posteriors, finding that a modification of the normalization of the stellar-mass-halo-mass relation resulted in shifted posteriors with physical interpretations. In our case, the limited flexibility of the SC-SAM prevents the stellar feedback parameters from compensating for this discrepancy. Greater flexibility in the model, such as varying additional feedback parameters, could enable better fitting, resulting in more realistic posterior values.

Pandya et al.~\cite{pandya2026introducingsapphirehybridphysicsinformed} also find that the magnitude of the measured error bars impact posteriors, with precision on some parameters increasing significantly when error bars are reduced. This sensitivity demonstrates the impact of assumed errors on astrophysical parameter constraints, particularly in the stellar feedback parameters. Similarly, we find that inflating the error on the~\cite{10.1093/mnras/stx1172} SHMR shifts the stellar feedback parameters slightly toward their fiducial values, and again shifts $b_{\phi}$ distribution towards lower values.

We find smaller parameter shifts when conditioning on the SMF and $M_*-Z_*$. Despite these shifts in the galaxy formation posteriors, the resulting $b_{\phi}$ distributions for the DESI-like galaxy sample remain consistent across all observables, despite disagreements in the measured galaxy formation posteriors as summarized in Table~\ref{tab:DataAnalysisSummary}.

The $b_{\phi}$ distributions conditioned on the SMF and $M_*-Z_*$ are in good agreement between the fiducial and data-based analyses, with $|\Delta\mu_{b_{\phi}}|<|0.05|$. The SHMR-conditioned distribution shows a larger mean shift of $\Delta\mu_{b_{\phi}}=0.063$; given the narrow posterior width ($\sigma_{b_{\phi}}=0.02$), this constitutes a statistically significant tension between the fiducial and data-based results, attributable to the shift in $\rm{A_{AGN}}$ driven by SC-SAM model deficiencies at high halo masses. All analyses consistently prefer lower $b_{\phi}$ values than the unweighted distribution.

We obtain fractional reductions in $\sigma_{b_{\phi}}$ of $88.4\%\ \textrm{and}\ 97.5\%$ when conditioning on the SMF and SHMR respectively, consistent with the simulation-based validation. The $M_*-Z_*$-conditioned distribution yields a $33.8\%$ reduction. 

As in Sec.~\ref{subsec:fidConst}, we quantify the information gain from the weighted or conditioned distribution $P$ relative to the unweighted distribution $Q$ using the K-L divergence, measuring values between $0.2\ \text{and}\ 3.16$ for the MCMC-conditioned distributions, with the SMF and SHMR values marginally exceeding those from the simulation-based analyses --- likely reflecting the larger posterior shifts induced by real data rather than genuine improvements in constraining power. The $\chi^2$-weighted distributions again provide negligible information gain (K-L values of $0$--$0.05)$ --- even when our galaxy model is known to not be flexible enough to fit all observables simultaneously, the information from conditioning on observation still considerably narrows the prior.

\begin{table*}
    \centering
    \begin{tabular*}{ \linewidth}{@{\extracolsep{\fill}} c|c|ccc|ccc}
        \multicolumn{8}{c}{\rule{0pt}{1cm}{\text{$b_{\phi}$ Priors Conditioned on Observed Data}}\rule[-1ex]{0pt}{1cm}}  \\
        \hline
         \rule{0pt}{0.7cm}Metric & $\langle b_{\phi}\rangle_{\rm{SU}}$ & \multicolumn{3}{c|}{$\langle b_{\phi}\rangle_{{\chi^2}_{\{Obs\}}}$} & \multicolumn{3}{c}{$p\left(b_{\phi} |\{Obs\}\right)$} \\
         &  & SMF & SHMR & $M_*-Z_*$ & SMF & $\text{SHMR}^{\dagger}$ & $M_*-Z_*$\rule{0pt}{0.7cm} \\
         \hline \hline
         $\langle b_{\phi}\rangle $ & $1.43$ & $1.20$ & $1.41$ & $1.39$ & $1.15$ & $1.25$ & $1.15$ \\
         $\sigma_{b_{\phi}} $ & $0.69$ & $0.48$ & $0.58$ & $0.62$ & $0.08$ & $0.02$ & $0.45$ \\
         $\left(\langle b_{\phi}\rangle - \langle b_{\phi}\rangle_{\rm{ref}}\right)/\sigma_{\rm{sym}}$ & --- & $-0.21$ & $-0.36$ & $-0.01$ & $-0.57$ & $-0.37$ & $-0.49$\\
         $\Delta\sigma_{b_{\phi}}/\sigma_{b_{\phi},\rm{ref}}$ & --- & $28.9\%$ & $16.0\%$ & $10.1\%$ & $88.9\%$ & $97.5\%$ & $33.8\%$\\
         K-L Divergence & --- & $0.19$ &$0.03$ & $0.06$ & $1.62$ & $3.16$ & $0.20$\\
         \hline
    \end{tabular*}
    \caption{Same as Table~\ref{tab:fidAnalysisSummary}, but for analyses conditioning on observational data vectors (Sec.~\ref{sec:ObsDataVec}). All conventions and definitions follow those of Table~\ref{tab:fidAnalysisSummary}.}\label{tab:DataAnalysisSummary}
\end{table*}

\section{Conclusion}\label{sec:Conclusion}

The increase in volume of current and forthcoming spectroscopic surveys offers the opportunity to improve upon CMB-based constraints on $f^{\rm{loc}}_{\rm{NL}}$, potentially discriminating between classes of inflationary models. Progress is limited by the exact degeneracy between $f^{\rm{loc}}_{\rm{NL}}$ and the bias parameter $b_{\phi}$ at the level of the galaxy two-point function. Current LSS analyses resolve this degeneracy through the universality relation, expressing $b_{\phi}$ as a function of $b_1$; however, this relation breaks down for physically motivated galaxy selections~\cite{Barreira_2022_bphi, Barreira_2020}, and the resulting systematic uncertainties in $b_{\phi}$ propagate directly into biases and precision losses in the inferred $f^{\rm{loc}}_{\rm{NL}}$.

We present a framework for constructing physically motivated, observation-conditioned priors on $b_{\phi}$ by marginalizing over galaxy formation uncertainties. We train Gaussian Process Emulators on the CAMELS-SAM suite to predict galaxy formation observables across the parameter space, use MCMC sampling to obtain the posterior $p\left(\boldsymbol{\theta}_{\rm{gf}}|\{Obs\}\right)$, and use a radial basis function interpolator trained on 50 separate universe simulations to map each posterior sample to a $b_{\phi}$ prediction, producing the marginalized prior $p\left(b_{\phi}|\{Obs\}\right)$. We target a galaxy sample replicating the DESI ELG selection.

When validated against observables from the fiducial CAMELS-SAM simulation, we find reductions in $\sigma_{b_{\phi}}$ of $88.4\%$ and $96.7\%$ when conditioning on the SMF and SHMR respectively, relative to the unweighted prior. When applied to observational data, we obtain reductions in $\sigma_{b_{\phi}}$ of $88.4\%$ and $97.5\%$ for the same observables. The resulting $b_{\phi}$ posteriors remain mutually consistent across all three observables despite shifts in the galaxy formation parameter posteriors --- most notably in $\rm{A_{AGN}}$ when conditioning on the SHMR --- driven by known discrepancies between the fiducial SC-SAM model and observational data at high halo masses. These discrepancies highlight the importance of marginalizing over uncertainties across multiple galaxy formation models, rather than within a single framework.

Future analyses should investigate galaxy observables that are robust to inter-model discrepancies. Expanded simulation suites would also enable the use of more flexible inference methods, such as neural density estimation, for constructing $b_{\phi}$ priors applicable to next-generation $f^{\rm{loc}}_{\rm{NL}}$ analyses.

\acknowledgments
We acknowledge Nick Kokron, Laurence Gong, and Joe Adamo for detailed comments on the manuscript. We thank Rachel Somerville, Shy Genel, Sebastian Wagner-Care\~{n}a, and Oliver Philcox for helpful discussions. AM and EK are grateful to the Flatiron Institute and to the CCA for their hospitality during the completion of this work. The CAMELS-SAM and the separate-universe simulations used here were run with the supercomputing resources of the Flatiron Institute, funded by the Simons Foundation.

\bibliographystyle{abbrvnat}
\setcitestyle{maxnames=3}
\bibliography{bibliography} 

\end{document}